\documentclass[
 reprint,
 superscriptaddress,
 amsmath,amssymb,
 aps,
 prb,
]{revtex4-2}

\usepackage{graphicx}
\usepackage{dcolumn}
\usepackage{bm}
\usepackage{color}
\usepackage{soul}
\graphicspath{{./}{./images/}}

\usepackage[colorlinks=true,      
    linkcolor=blue,      
    citecolor=blue,       
    urlcolor=blue,        
    breaklinks=true      
    ]{hyperref}
\begin{document}

\preprint{APS/123-QED}

\title{Magic transition in monitored free fermion dynamics}

\author{Cheng Wang}
\affiliation{School of Physics, Peking University, Beijing 100871, China}
\author{Zhi-Cheng Yang}
\affiliation{School of Physics, Peking University, Beijing 100871, China}
\affiliation{Center for High Energy Physics, Peking University, Beijing 100871, China}
\author{Tianci Zhou}
\affiliation{Department of Physics, Virginia Tech, Blacksburg, Virginia 24061, USA}
\author{Xiao Chen}
\email{chenaad@bc.edu}
\affiliation{Department of Physics, Boston College, Chestnut Hill, MA 02467, USA}

\date{\today}

\begin{abstract}

We investigate magic and its connection to entanglement in 1+1 dimensional random free fermion circuits, with a focus on hybrid free fermion dynamics that can exhibit an entanglement phase transition. To quantify magic, we use the Stabilizer R\'enyi Entropy (SRE), which we compute numerically via a perfect sampling algorithm. We show that although the SRE remains extensive as the system transitions from a critical phase to an area-law (disentangled) phase, the structure of magic itself undergoes a delocalization phase transition. This transition is characterized using the bipartite stabilizer mutual information, which exhibits the same scaling behavior as entanglement entropy: logarithmic scaling in the critical phase and a finite constant in the area-law phase. Additionally, we explore the dynamics of SRE. While the 
total SRE becomes extensive in 
$O(1)$ time, we find that in the critical phase, the relaxation time to the steady-state value is parameterically longer than that in generic random circuits. The relaxation follows a universal form, with a relaxation time that grows linearly with the system size, providing further evidence for the critical nature of the phase.

\end{abstract}

\maketitle

\section{Introduction}

Entanglement has long been a central focus of quantum many-body physics. It characterizes the non-locality of quantum states and encapsulates the long-range correlations within a system. Researchers have identified entanglement as a fundamental intrinsic property, probing many emergent phenomena, such as quantum phases of matter \cite{amico2008entanglementmanybody}, quantum phase transitions \cite{Osterloh2002}, quantum chaos \cite{srednicki1994chaosquantumthermalization,Hosur2016chaosquantumchannel}, and quantum dynamics \cite{nahum2017entanglementgrowth,Nahum2018operatorspreading}. Beyond its significance in understanding the quantum nature of many-body systems, entanglement also plays a critical role as a quantum resource in the context of quantum computation \cite{Nielsen_Chuang_2010}. It determines the depth required to prepare a specific quantum state if only local operations are allowed. The deeper the circuit, the more challenging and costly it becomes to prepare the state on near-term quantum devices.

However, an aspect often overlooked is the cost of individual quantum operations. Some quantum operations, such as Clifford unitaries, are relatively inexpensive to implement both on quantum devices and in classical simulations. In contrast, other operations, commonly referred to as magic or non-Clifford operations, are significantly more resource-intensive. To accurately characterize the total complexity of a quantum circuit, it is essential to account for both the total number of gates and the cost of each gate. As a result, non-Cliffordness, or more colloquially, magic,  is as important a quantum resource as entanglement.

In recent years, various measures have been proposed to quantify magic, among which the Stabilizer Rényi Entropy (SRE) stands out as a computationally tractable measure \cite{leone2022stabilizerrenyientropy,Haug2023stabilizerentropies}. The SRE computes the entropy of a quantum state’s density operator expressed in the Pauli string basis and estimates the deviation from a stabilizer state. It is a magic monotone under resource theory (for R\'enyi index $n > 2$), meaning that it only counts the non-Clifford operations  \cite{leone2022stabilizerrenyientropy}. 

While both magic and entanglement are recognized as quantum resources, our understanding of the magic's role in many-body dynamics and its connection to entanglement remains limited. The gap arises from two key challenges. First, even computable measures of magic, such as SRE, are more challenging to calculate than entanglement. The limited computable examples of magic in many-body systems significantly restrict our exploration of its properties.  Consequently, most existing research of magic has focused on low-entanglement systems, where the quantum states can be at least approximately constructed efficiently. Although analytic attempts have been made in highly entangled systems, they either use different magic measures \cite{zhang2024quantummagicdynamicsrandom} or are restricted to a specific set of Hamiltonian evolutions \cite{Lopez2024sreXXZmodel}. Second, SRE often has a mundane, extensive scaling in many unitary and hybrid evolutions, including ones presented in this work. The extensive scaling has uninteresting local contributions, which shadow the non-local part that intermingles with the entanglement and features of the quantum phases. For instance, a random product state, despite having zero entanglement, possesses extensive SRE. Therefore, intriguing quantum phases that manifest through non-local correlations or specific entanglement structures may not be readily discernible through the direct scaling behavior of the total SRE.

To address these challenges, we investigate free fermion circuits where quantum states can have large entanglement while still being efficiently simulable using classical algorithms. We focus on hybrid quantum circuits involving both unitary free fermion gates and non-unitary quadratic measurement gates. These models host an entanglement phase transition as a function of the measurement rate: a critical phase with logarithmic entanglement scaling at low measurement rates transitions into an area-law phase as the measurement rate increases \cite{bastianello2018quasiparticlesspreadingentanglement, chen2020conformalfreefermion, Fava2023nonlinearsigma, fava2024monitoredfermionsU(1), guo2025fieldtheorymonitoredinteracting,Alberton_2021,Zhang_2021,Poboiko_2023,Cao_2019,Turkeshi_2021,li2018quantumzeno,Skinner_2019}.
By our construction, the state under hybrid evolution remains a Gaussian free fermion state. It's well known that its entanglement can be computed efficiently, and so can the SRE through a perfect sampling method \cite{collura2024paulisamplingfermion, tarabunga2024magictransitionmeasurementonlycircuits}. This sampling approach enables us to compute SRE with up to one hundred sites.

In the steady state, SRE exhibits volume-law scaling in both the critical and area-law phases. However, we argue that despite the extensive SRE, the underlying structure of the magic is significantly different. In the area-law phase, magic are localized to individual qubits and its short range entangled neighbors, while in the critical phase, entanglement can support long-range magic that cannot be removed by local unitaries. 

To quantify such non-locality of the magic, we introduce the stabilizer mutual information (SMI), focusing in particular on the bipartite SMI (BSMI) of the quantum state. We show that BSMI exhibits scaling behavior analogous to that of entanglement entropy in the measurement-induced phase transition. That is, it scales logarithmically in system size in the critical phase, and approaches an ${O}(1)$ constant in the area-law phase. The coincidence of the scaling behaviors of SMI and entanglement entropy suggests that along side the entanglement phase transition, there is a concurrent delocalization transition of magic that is effectively captured by the SMI. 

In both pure random unitary dynamics and the critical phase, we find that the SRE undergoes a slow relaxation. With low measurement rate, the system enters a critical phase, and the growth of SRE collapses as a universal form of $t/L$, which entails a linear saturation time. The scaling function features the universality of the critical phase \cite{chen2020conformalfreefermion,Li_2021}. In the trivial area-law phase, such scaling behavior is absent. 

The remainder of the paper is organized as follows. In Sec.~\ref{sec:sre_smi}, we introduce the non-unitary random free fermion circuit and describe the algorithm used to efficiently compute the SRE and SMI. A more detailed discussion of the SMI on random states is provided in App.~\ref{app:detail_def_SMI}. In Sec.~\ref{sec:mi_sre_mipt}, we present our numerical results on simulating non-unitary free-fermion circuits. Sec.~\ref{sec:steady} focuses on the steady-state properties, while Sec.~\ref{sec:dyn} discusses the dynamics of SRE and SMI. Both quantities are used to characterize the critical phase. We conclude by summarizing our results in Sec.~\ref{sec:con}.

\section{Stabilizer R\'enyi Entropy and Stabilizer Mutual Information}

\label{sec:sre_smi}

Quantum magic, or non-stabilizerness is a crucial resource of achieving universal quantum computing and is formally studied in the framework of resource theory~\cite{Veitch2014resourcetheorystabilizer}. In this context, stabilizer states are designated as free states, which can be prepared and manipulated using the free Clifford operations. A magic measure $M(\rho)$ quantifies the amount of non-stabilizerness in a quantum state $\rho$. Among the magic measures proposed, the stabilizer R\'enyi entropy is the most tractable ones \cite{leone2022stabilizerrenyientropy}, at least for certain classes of states such as the matrix product states.

The SRE is defined by quantifying the ``spread'' of the quantum state $\rho$ among the Pauli string basis. For an $n$-qubit system, let $\{ P_i \}_{i=0}^{ D^2 - 1} $ be the set of $N$-qubit Pauli strings, where $D = 2^N$ is the Hilbert space dimension.  The weight for the normalized Pauli operator $\frac{P_i}{\sqrt{D}}$ is $w_i = |\frac{1}{\sqrt{D}} \text{Tr}(\rho P_i)|^2$, which can be viewed as a probability distribution for a pure quantum state $\rho$. The SRE is the R\'enyi-$\alpha$ entropy of  this distribution shifted by $\ln D$ so that stabilizer states have zero SRE. 

The intuition of SRE can be understood as follows. It is well known that the density matrix of a pure stabilizer state can be written as an equal weight superposition of all its stabilizers. Thus for a stabilizer state, there are only $D$ out of $D^2$ possible values of $\text{Tr}( \rho P_k)^2$ that are non-zero, leading to a sparse distribution of $w_i$. For example, a single qubit stabilizer state $|0\rangle \langle 0| = \frac{1}{2} ( I + Z)$ only have support on $Z$ and $I$. In contrast, a non-stabilizer (magic) state can generally have non-zero values on a much larger set of Pauli operators, possibly spreading over the entire $D^2$ Pauli strings. Therefore the degree of the non-stabilizerness is captured by the entropy of the Pauli weight.

Formally, the SRE is defined as (\cite{leone2022stabilizerrenyientropy})
\begin{equation}
    M_{\alpha}(\rho) = \frac{1}{1-\alpha} \log \bigg[ \frac{\sum_P \mathrm{Tr}^{2\alpha}(\rho P)}{\sum_P \mathrm{Tr}^{2}(\rho P)} \bigg].
\end{equation}
In particular, $M_1(\rho)$ is obtained through the analytic continuation $\alpha \rightarrow 1$: 
\begin{equation}
    M_{1}(\rho) = -\frac{\sum_P \mathrm{Tr}^{2}(\rho P) \log \big[ \mathrm{Tr}^{2}(\rho P) \big]}{\sum_P \mathrm{Tr}^{2}(\rho P)}.
\end{equation}
Notably, the definition extends to the mixed states. And they all satisfy several key properties required by the resource theory: 
(1) \textbf{Faithfulness}: SRE is zero for stabilizer states and strictly positive for non-stabilizer states.  
(2) \textbf{Stability}: SRE remains invariant under free operations such as Clifford gates.  
(3) \textbf{Additivity}: SRE is additive under the tensor product of two states, i.e.,  
\begin{equation}
    M_{\alpha}(\rho_A \otimes \rho_B) = M_{\alpha}(\rho_A) + M_{\alpha}(\rho_B).
\end{equation}

The interplay between entanglement and magic is an active research topic \cite{zhang2024quantummagicdynamicsrandom, hou2025stabilizerentanglementmagichighway, viscardi2025interplayentanglementstructuresstabilizer}. On one hand, a quantum state can possess extensive magic even with low entanglement. For example, a tensor product of single qubit magic state, such as the $T$ state, contains magic that scales with the system size, yet the magic is completely localized to individual qubits. On the other hand, a zero magic state -- stabilizer state can be highly entangled, or even maximally entangled, such as a collection of Bell pairs. Of particular interest is the magic that is intrinsically delocalized, intertwining with the non-local entanglement between the subregions of the system. Such non-local magic cannot be eliminated by local unitary operators confined in the subregions. Quantifying this type of non-local magic has primitive results, e.g., through magic measures after optimization over local unitaries~\cite{ cao_gravitational_2025, hou2025stabilizerentanglementmagichighway}.

In this work, we explore an alternative diagnostic, the stabilizer mutual information (SMI) defined in the same fashion as the conventional mutual information for entanglement: $M_{\alpha}^A + M_{\alpha}^{\bar{A}} - M_{\alpha}^{A\cup \bar{A}}$. However due to lack of a strict subadditivity structure from $M_{\alpha}$, the quantity thus defined does not guarantee a definite sign. One might na\"ively enforce the sign by $|M_{\alpha}^A + M_{\alpha}^{\bar{A}} - M_{\alpha}^{A\cup \bar{A}}|$. However, based on the results of random states and the observations of the circuits studied in this work (App.~\ref{app:detail_def_SMI}), we adopt the following sign convention 
\begin{equation}
\label{eq:SMI}
    I_{\alpha} \equiv
    \begin{cases}
        M_{\alpha}^A + M_{\alpha}^{\bar{A}} - M_{\alpha}^{A\cup \bar{A}}, & \alpha \leq 1 \\
        - M_{\alpha}^A - M_{\alpha}^{\bar{A}} + M_{\alpha}^{A\cup \bar{A}}, & \alpha \geq 2
    \end{cases}.
\end{equation}
which ensures a non-negative $I_\alpha$ in the relevant regimes. This quantity can characterize the non-local magic supported between $A$ and $\bar{A}$. 

\subsection{Free fermion dynamics and the sampling algorithm of SRE}
\label{subsec:sampling_algorithm}

Although SRE can be written in a relatively simple form for quantifying magic, computing the SRE in quantum many-body systems is generally a challenging task. This difficulty arises not only from the complexity of computing the SRE itself but also from the lack of effective techniques for characterizing highly entangled states with extensive magic when the stabilizer formalism loses its applicability. Recent advances have introduced methods based on Pauli sampling, enabling efficient estimation of SRE in systems with low entanglement, such as matrix product states or the ground states of certain Hamiltonians \cite{Haug2023quantifyingmagicmps, tarabunga2023magicpaulisampling, lami2023paulisamplingMPS, liu2024paulisamplingspin}. However, these approaches remain limited to low-entanglement states and are insufficient for capturing the relationship between quantum magic and entanglement in more general settings. Simulating quantum systems with both extensive entanglement and extensive magic poses significant computational challenges. To make progress, we focus on free fermionic systems described by Gaussian states. This approach is advantageous for two reasons: (1) fermionic Gaussian states are entirely characterized by their $2N \times 2N$ Majorana covariance matrix (for an $N$-mode system). Dynamics and measurements by fermion bilinear operators preserve this Gaussian nature, allowing for simulations with resources polynomial in $N$. (2) For Gaussian states, the expectation values $\mathrm{Tr}(\rho P_k)$ of Pauli strings $P_k$ (required for SRE calculation) can be computed efficiently. This is achieved by first mapping Pauli strings to Majorana strings---products of Majorana fermion operators---via the Jordan-Wigner transformation, and then evaluating their expectation values using Wick's theorem. 

Below, we review and establish notation for fermionic Gaussian states via their Majorana operators and covariance matrix. We then discuss the method to sample the Majorana strings and compute the SRE.

We begin with the Jordan-Wigner transformation, which maps $N$ spin-$1/2$ operators on sites $i=0, \dots, N-1$ to $2N$ Majorana fermion operators $\gamma_{\mu}$ with $\mu = 0, \dots, 2N-1$:
\begin{equation}
\label{eq:JW_transformation}
    \gamma_{2i} = \left(\prod_{j<i} Z_j\right) X_i, \quad \quad
    \gamma_{2i+1} = \left(\prod_{j<i} Z_j\right) Y_i.
\end{equation}
They satisfy the fermion algebra $\{\gamma_{\mu}, \gamma_{\nu}\} = 2\delta_{\mu\nu}$.

A fermionic Gaussian state $\rho$ is fully characterized by its Majorana covariance matrix $\Gamma(\rho)$:
\begin{equation}
    \Gamma_{\mu \nu} = -\frac{i}{2} \langle [\gamma_{\mu},\gamma_{\nu}]\rangle,
\end{equation}
which is real and antisymmetric. 

For numerical implementations, notice that a fermionic Gaussian state can be fully characterized by $N$ fermionic operators -- namely $d_1,~d_2, \cdots,~d_N$ -- that annihilate the state. These operators can be expressed as linear combinations of $2N$ Majorana operators as
\begin{equation}
    d_{\mu}=\sum_{\kappa=0}^{2N-1} \alpha_{\mu\kappa} \gamma_{\kappa}.
\end{equation}
If we define the column vector $\bm{\gamma}$ and $\bm{d}$ as,
\begin{equation}
    \bm{\gamma}=\begin{bmatrix}
        \gamma_1\\
        \cdots \\
        \gamma_{2N}
    \end{bmatrix},~\bm{d}=\begin{bmatrix}
        d_1\\
        \cdots \\
        d_{N}
    \end{bmatrix}.
\end{equation}
The annihilation operators can be written as $\bm{d}=\alpha \bm{\gamma}$

To ensure that the annihilation operators $d_1,\cdots, d_N$ satisfy the anticommutativity, $\{d_{\mu}^{\dagger}, d_{\nu}\}=\delta_{\mu\nu}$, the $N\times 2N$ matrix $\alpha_{\mu\nu}$ has to follow the orthonormality condition, 
\begin{equation}
    (\alpha\alpha^{\dagger})_{\mu\nu}=\sum_{\kappa=0}^{2N-1}\alpha_{\mu\kappa}\alpha_{\nu\kappa}^{*} =\frac{1}{2} \delta_{\mu\nu}.
\end{equation}
Correspondingly, the Majorana covariance matrix can also be expressed as,
\begin{equation}
    \Gamma_{\mu \nu}=-2i [(\alpha^{\dagger}\alpha)_{\mu\nu}-(\alpha^{\dagger}\alpha)_{\nu\mu}]=-2i\sum_{\kappa=0}^{2N-1}(\alpha_{\kappa\mu}^{*}\alpha_{\kappa\nu}-\alpha_{\kappa\nu}^{*}\alpha_{\kappa\mu}). 
\end{equation}

For unitary dynamics, the evolution is controlled by a quadratic Hamiltonian $U_{\tau}=\exp(-i\tau H_1)$ with a general form $H_1 = \frac{i}{4} \sum_{\mu, \nu=0}^{2N-1} h_{\mu\nu} \gamma_{\mu}\gamma_{\nu}$, where $h$ is a real, antisymmetric matrix. The annihilation operators $d_{\mu}$ transform accordingly under unitary evolution,
\begin{equation}
\begin{aligned}
|\psi\rangle &\to U_{\tau}|\psi\rangle,\\
    d_{\mu} &\to U_{\tau}d_{\mu}U_{\tau}^{\dagger}.
\end{aligned}
\end{equation}

The matrix $\alpha$ that governs the annihilation operators also changes linearly,
\begin{equation}
    \alpha_{\mu\nu}\to(\alpha O_\tau)_{\mu\nu}=\sum_{\kappa=0}^{2N-1}\alpha_{\mu\kappa}O_{\tau,\kappa\nu},
\end{equation}
where $O_\tau= \exp(-\tau h)$ is a real orthogonal matrix in group $\mathrm{SO}(2N)$. The orthonormality condition for $\alpha$ is automatically satisfied after unitary evolution.

As for non-unitary dynamics, such as weak measurements, simulations can also be efficiently performed using fermionic Gaussian states. In this case, the state evolves under imaginary-time evolution given by a Kraus operator $K_{\pm\beta} = \exp(\pm \beta H_2)$, where the Hamiltonian is given by $H_2 = \sum_{\mu\nu} -i\lambda_{\mu\nu} \gamma_{\mu} \gamma_{\nu}$ and $\lambda$ is also a real and antisymmetric matrix. For projective or weak measurements, the signs are determined according to the Born rule, based on the measurement outcomes. In contrast, for forced measurements or projections, these signs are fixed beforehand.

Under non-unitary evolution, the annihilation operators change similarly as the unitary case \cite{Ravindranath_2025,Ravindranath_2023}, 
\begin{equation}
    \begin{aligned}
        |\psi\rangle&\to\frac{K_{\pm \beta}|\psi\rangle}{\langle\psi|K_{\pm \beta}^2|\psi\rangle}, \\
        d_{\mu} &\to \tilde{d_\mu}=K_{\pm \beta} d_\mu K_{\mp \beta}.
    \end{aligned}
\end{equation}
and the matrix $\alpha$ will also follow a linear transformation, given by,
\begin{equation}
    \alpha_{\mu\nu}\to\tilde{\alpha}=(\alpha A_{\pm \beta})_{\mu\nu}=\sum_{\kappa=0}^{2N-1}\alpha_{\mu\kappa}A_{\pm\beta,\kappa\nu},
\end{equation}
where $A_{\pm \beta}=\exp(\mp i\beta\lambda)$. It should be noticed that here $A_{\pm \beta}$ is not a real orthogonal matrix as $O_{\tau}$ before and the orthonormality of  matrix $\alpha$ will be broken after the non-unitary evolution. However, the annihilation operators are still independent from each other and we only need to perform a Gram-Schmidt process to obtain an orthonormal matrix $\alpha$ for the state. Such a simple algorithm for computing $\alpha$ provides us enough information about quantum evolution of the Gaussian states and makes it possible for us to efficiently compute the physical quantities we want, such as EE and SRE.

The SRE quantifies how a density matrix $\rho$ is distributed across the basis of Pauli strings. It is formally the R\'enyi entropy of the probability distribution generated by the squared expectation values of these operators. The Jordan-Wigner transformation establishes a one-to-one correspondence between Pauli strings and Majorana strings, which differ only by a phase factor. Consequently, the set of squared expectation values for Majorana strings is a permutation of that for Pauli strings, meaning both bases yield the same entropy. We can therefore compute the SRE in the Majorana basis. Representing a Majorana string operator as $\gamma_{\mathbf{x}} = \gamma_0^{x_0} \gamma_1^{x_1} \cdots \gamma_{2L-1}^{x_{2L-1}}$ using a binary vector $\mathbf{x} = (x_0, x_1, \dots, x_{2L-1})$, the SRE is written as:
\begin{equation}
    M_{\alpha}(\rho) = \frac{1}{1-\alpha} \log \bigg[ \frac{\sum_{\mathbf{x}} |\mathrm{Tr}(\rho \gamma_{\mathbf{x}})|^{2\alpha}}{\sum_{\mathbf{x}} |\mathrm{Tr}(\rho \gamma_{\mathbf{x}})|^{2}} \bigg],
\end{equation}
where the summation is over all $2^{2L}$ distinct Majorana strings indexed by $\mathbf{x}$. For Gaussian states, the expectation value of each Majorana operator can be efficiently computed from the antisymmetric covariance matrix $\Gamma$ using Wick's theorem:
\begin{equation}
    \mathrm{Tr}(\rho \gamma_{\mathbf{x}}) = i^{|\mathbf{x}|/2} \mathrm{Pf}[\Gamma|_{\mathbf{x}}],
\end{equation}
where $\Gamma|_{\mathbf{x}}$ is the principal minor of $\Gamma$ containing the rows and columns indexed by the set $\{i | x_i=1\}$, and $|\mathbf{x}|$ is the number of Majorana operators in the string (The Pfaffian vanishes when $|{\bf x}|$ is odd).

While Wick's theorem and the corresponding Pfaffian formula allow each expectation value $\mathrm{Tr}(\rho \gamma_{\mathbf{x}})$ to be computed in polynomial time, the SRE expression involves a sum over all $4^L$ Majorana strings. Directly enumerating these terms is computationally intractable for large systems. Instead, we view the SRE calculation as a statistical sampling problem. We treat the Majorana strings as an ensemble governed by the probability distribution:
\begin{equation}
    \pi_{\rho}(\mathbf{x}) = \frac{|\mathrm{Tr}(\rho \gamma_{\mathbf{x}})|^2}{\sum_{\mathbf{y}} |\mathrm{Tr}(\rho \gamma_{\mathbf{y}})|^2}.
\end{equation}
By drawing samples $\mathbf{x}$ from the distribution $\pi_{\rho}(\mathbf{x})$, the SRE is approximated as an ensemble average:
\begin{equation}
    M_{\alpha}(\rho) \approx \frac{1}{1-\alpha}\log\left[\left\langle |\mathrm{Tr}(\rho \gamma_{\mathbf{x}})|^{2\alpha-2}\right\rangle_{\pi_\rho} \right],
\end{equation}
where $\langle \cdot \rangle_{\pi_\rho}$ denotes the average over samples drawn from $\pi_{\rho}(\mathbf{x})$. In the case of $\alpha=1$, this simplifies to:
\begin{equation}
    M_1(\rho) \approx -\left\langle \log\left[ |\mathrm{Tr}(\rho \gamma_{\mathbf{x}})|^2\right] \right\rangle_{\pi_\rho}.
\end{equation}
Consequently, the problem of computing the SRE is transformed into sampling the Majorana strings according to the distribution $\pi_{\rho}(\mathbf{x})$.

A straightforward approach to sampling $\pi_\rho(\mathbf{x})$ is to employ a standard Markov chain Monte Carlo method, such as the Metropolis-Hastings algorithm. While this approach is generally applicable, it requires a thermalization time to ensure the samples are close enough to the target distribution. However, a more efficient ``perfect sampler" can be constructed, as used in Ref.~\cite{collura2024paulisamplingfermion}. This method, analogous to techniques for sampling Paulis from matrix product states (MPS), uses the chain rule of probability to sample the binary vector $\mathbf{x}$ sequentially:
\begin{equation}
    \pi_{\rho}(\mathbf{x})=\pi_{\rho}(x_0)\pi_{\rho}(x_1|x_0)\cdots\pi_{\rho}(x_{2L-1}|x_{0}, \dots, x_{2L-2}). 
\end{equation}
Each conditional probability is then expressed as a ratio of marginal probabilities:
\begin{equation}
    \pi_{\rho}(x_{\mu} | x_0, \dots, x_{\mu-1}) = \frac{\pi_{\rho}(x_0, \dots, x_{\mu})}{\pi_{\rho}(x_0, \dots, x_{\mu-1})}. 
\end{equation}
The key observation is that these marginals can be computed efficiently. Recalling that $|\mathrm{Tr}(\rho \gamma_{\mathbf{x}})|^2 = \det(\Gamma|_{\mathbf{x}})$ for Gaussian states, the problem reduces to summing determinants of principal minors. A well-known identity states that the sum over all principal minors of a matrix is given by:
\begin{equation}
  \sum_{\mathbf{x}} \det( \Gamma|_{\mathbf{x}} )  = \det ( \mathbb{I} + \Gamma ).
\end{equation}
More generally, the marginals required for the chain rule can be expressed as:
\begin{equation}
    \sum_{x_{\mu+1}, \dots, x_{2L-1}} \det( \Gamma|_{\mathbf{x}} ) = \det\left[ (\mathbb{I}_{[\mu+1, 2L-1]}+\Gamma)\big|_{(x_0\cdots x_{\mu},1\cdots1)}\right].
\end{equation}
This leads to a closed-form expression for the marginal distribution:
\begin{equation}
    \pi_{\rho}(x_0, \dots, x_{\mu}) = \frac{\det\left[ (\mathbb{I}_{[\mu+1, 2L-1]}+\Gamma)\big|_{(x_0\cdots x_{\mu},1\cdots1)}\right]}{\det( \mathbb{I} + \Gamma)}.
\end{equation}
Here, $\mathbb{I}_{[\mu+1, 2L-1]}$ denotes a diagonal matrix with ones in the rows and columns indexed by $\{\mu+1, \dots, 2L-1\}$ and zeros elsewhere.

The determinant of a $2L \times 2L$ matrix can be computed with a cost of $\mathcal{O}(L^3)$. Since generating a single Majorana string sample requires sequentially determining $2L$ bits, the total computational complexity for one sample is $\mathcal{O}(L^4)$. Consequently, this method enables the calculation of the SRE for free-fermion systems in polynomial time.

The efficient sampling scheme above can be directly applied to compute the SMI defined in Eq.~\ref{eq:SMI}, which involves terms like $M_{\alpha}(\rho_A)$, $M_{\alpha}(\rho_B)$, and $M_{\alpha}(\rho_{A \cup B})$. The term for the full system, $M_{\alpha}(\rho_{A \cup B})$, is the total SRE discussed above. The subsystem terms, such as $M_{\alpha}(\rho_A)$, are calculated by restricting the sampling to Majorana operators with support only within region $A$. Computationally, this is achieved by applying the same sampling methodology to the principal minors of $\Gamma$ corresponding only to the indices within the target region.

\section{Numerical results in free-fermion circuits}

\label{sec:mi_sre_mipt}

\subsection{Steady states of non-unitary random evolutions}
\label{sec:steady}
\begin{figure}
    \centering
    \includegraphics[width=\linewidth]{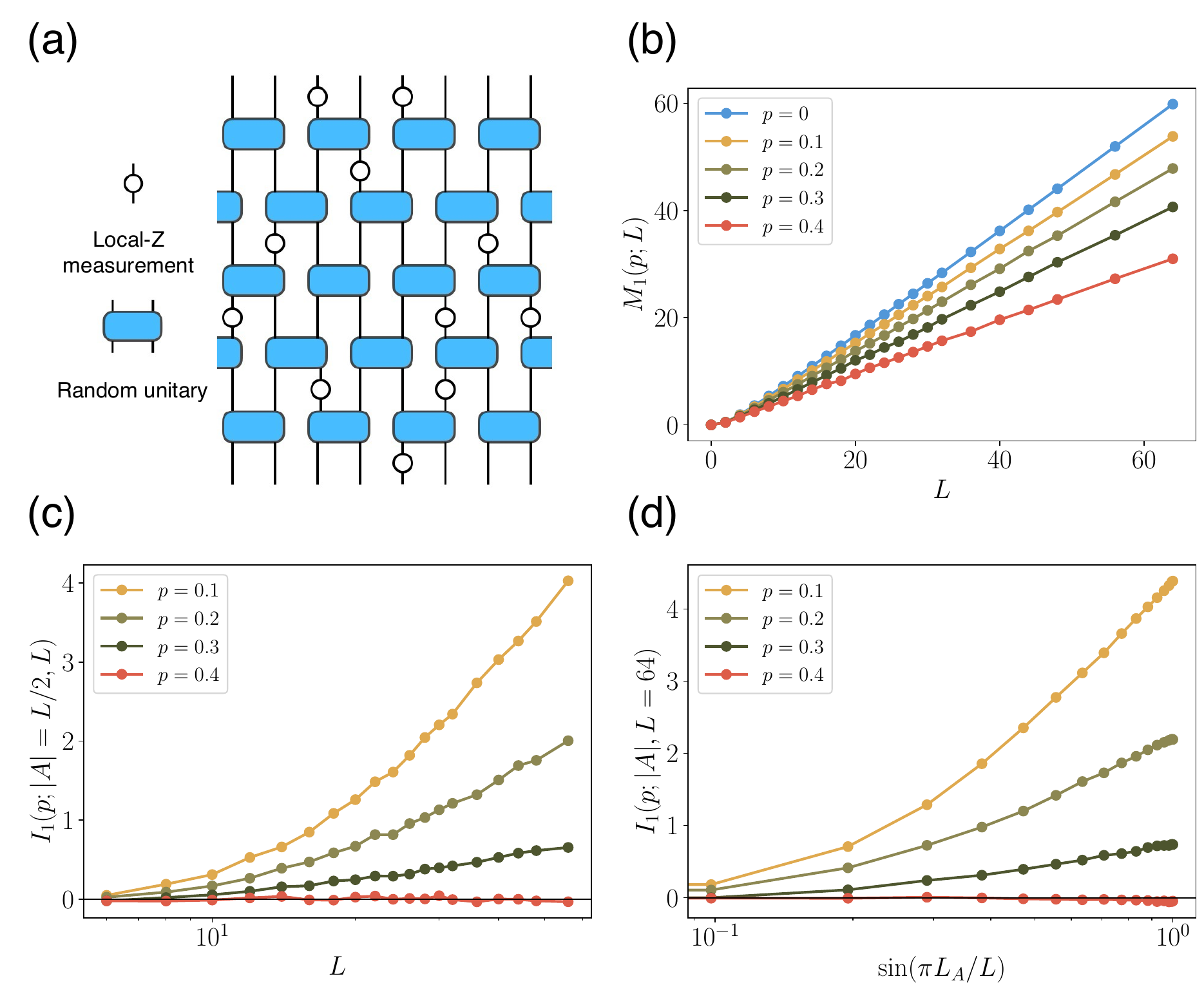}
    \caption{SRE and  BSMI of steady states generated by unitary free fermion circuits with random projective $Z$ measurements. (a) Structure of the 1+1D free fermion hybrid circuits. Local-$Z$ measurements are applied with probability $p$. (b) The total SRE of steady states for different $p$, all of which scale linearly with the system size, $M\sim L$. (c) The BSMI with a fixed subsystem fraction ($|A|/L = 1/2$) as a function of system size $L$. Without measurements ($p = 0$, not shown), the SMI follows a volume-law scaling. Finite measurement rates ($p>0$) leads to a logarithmic scaling, $I\sim \log L$. Above a critical measurement rate ($0.3 < p_c < 0.4$ the BSMI transitions to an area-law scaling, $I\sim {\rm const}$. (d) The BSMI for a fixed system size ($L = 64$) as a function of subsystem size $L_A$. We take $\sin(\pi L_A / L)$ as the $x$-axis to account for the periodic boundary conditions on a cylindrical geometry. As the subsystem size varies, the BSMI undergoes a transition from logarithmic scaling, $I\sim\log(\sin(\pi L_A / L))$, to an area-law scaling above the critical point.
}
    \label{fig:mi_system_size}
\end{figure}

We study magic in the hybrid free fermion circuits. In such 1+1D circuits, it is established that entanglement can exhibit a critical phase below a threshold measurement rate~\cite{Fava2023nonlinearsigma}. In this phase, the entanglement has logarithmic critical scaling, whereas for measurement rate beyond the threshold, the entanglement follows an area law. In this setup, we find that magic, quantified by the SRE, undergoes a ``localization'' transition as evidenced by the numerical results presented below. 

To specify the model, we consider a 1+1D free fermion circuit with mid-circuit projective measurements. The brickwork circuit consists of gates implementing free fermion unitary evolution and acting locally on neighboring sites in the Pauli basis.  The time evolution operator can be written as 
\begin{equation}
\begin{aligned}
    &U(t)=\prod_{\tau=1}^tU_{\rm odd}(\tau)U_{\rm even}(\tau)\\
    &U_{\rm odd/even}(\tau)=\bigotimes_{i\in {\rm odd/even}}U_{i,i+1}(\tau),
\end{aligned}
\end{equation}
which consists of layers with two-qubit gates acting on odd/even bonds. Each gate $U_{i,i+1}(\tau)=\exp(-{\rm i}H_{i,i+1}(\tau))$ is generated by a local Hamiltonian of the form,

\begin{equation}
    H_{i,i+1}(\tau)=\sum_{m,n\in\{2i,2i+1,2i+2,2i+3\},m\neq n}{\rm i}\kappa_{m,n}(\tau)\gamma_m\gamma_n,
    \label{Eq:unitary}
\end{equation}
where $\kappa_{m,n}(\tau)$ are sampled uniformly from $[0,\pi)$. In the Pauli operator basis, the Hamiltonian is a combination of $X_{i}X_{i+1}$, $X_{i}Y_{i+1}$, $Y_{i}X_{i+1}$, $Y_{i}Y_{i+1}$, $Z_{i}$ and $Z_{i+1}$, which contains up to 2-body interactions in the qubit language. Projective measurements are performed in the $Z$-basis. Both unitary gates and measurements preserve the free fermion Gaussian structure in all steps. The Gaussian nature allows an efficient computation of both entanglement and SRE, the latter through a sampling algorithm described in Sec.~\ref{subsec:sampling_algorithm}. 

We first note that entanglement and SRE are generated in fundamentally different ways. Entanglement can only be produced by gates that connect the two subsystems across a partition, and its growth rate is constrained by the locality of the interaction. Although projective measurements tend to destroy entanglement, their competition with unitary evolution can reach a balance in free fermion systems, which gives rise to a critical phase when the measurement rate is low. In contrast, quantum magic can be generated locally by any non-Clifford gate, even without multi-party interactions that produce entanglement. In our circuit, the unitary gates implementing free-fermion evolution are non-Clifford gates. As a result, there are order $\mathcal{O}(L)$ gates that can inject magic into the system at each time step. Projective measurements are not strong enough to completely suppress its growth. In Fig.~\ref{fig:mi_system_size}(b), we show the long time steady state value of SRE for the whole system as a function of the measurement rate $ p $. As $ p $ increases, the SRE decreases but still exhibits volume law scaling at low measurement rates, with no clear sign of a transition to an area law.

However, we note that at large measurement rates, the system enters an area-law phase for entanglement. A simplified picture is that the system becomes effectively fragmented into nearly disconnected regions. As a result, the observed volume law scaling of the SRE in this regime is largely due to local contributions from these isolated regions. In contrast, at low measurement rates, where entanglement exhibits critical scaling, the magic cannot be attributed to local contributions \textit{alone}. Instead, it contains a non-local part of magic, which can only be generated by interacting gates accompanying the growth of  entanglement. It is this non-local part of magic that probes the measurement induced criticality. 

We observe a magic ``localization'' transition in SMI defined in Eq.~\ref{eq:SMI}, which through its definition cancels the local contributions. In Fig.~\ref{fig:mi_system_size} (c), we plot the bipartite SMI (BSMI) as a function of system size for different measurement rates.
Specifically, when there is only unitary evolution and no measurement ($p = 0$), both steady state entanglement entropy and BSMI scale as a linear function of system size $L$ (not shown). When the measurement rate is low but nonzero, both the entanglement entropy and BSMI scale as $\log L$~\footnote{Strictly speaking, the entanglement entropy scales as $(\log L)^2$ and it is possible that the BSMI follows the same scaling~\cite{Fava2023nonlinearsigma}. However, this logarithmic correction is difficult to resolve numerically for both quantities.}. However, when the measurement rate exceeds the entanglement transition threshold $p_c$, the BSMI saturates to a finite constant, independent of system size. Such scaling is corroborated by Fig.~\ref{fig:mi_system_size} (d) when we vary the subsystem size $L_A$ and plot against the periodic variable $\sin( \pi L_A / L)$. The data suggest that the transition point of magic delocalization coincides with the entanglement transition, and entanglement in the critical phase is responsible for supporting non-local magic.

\begin{figure}
    \centering
    \includegraphics[width=\linewidth]{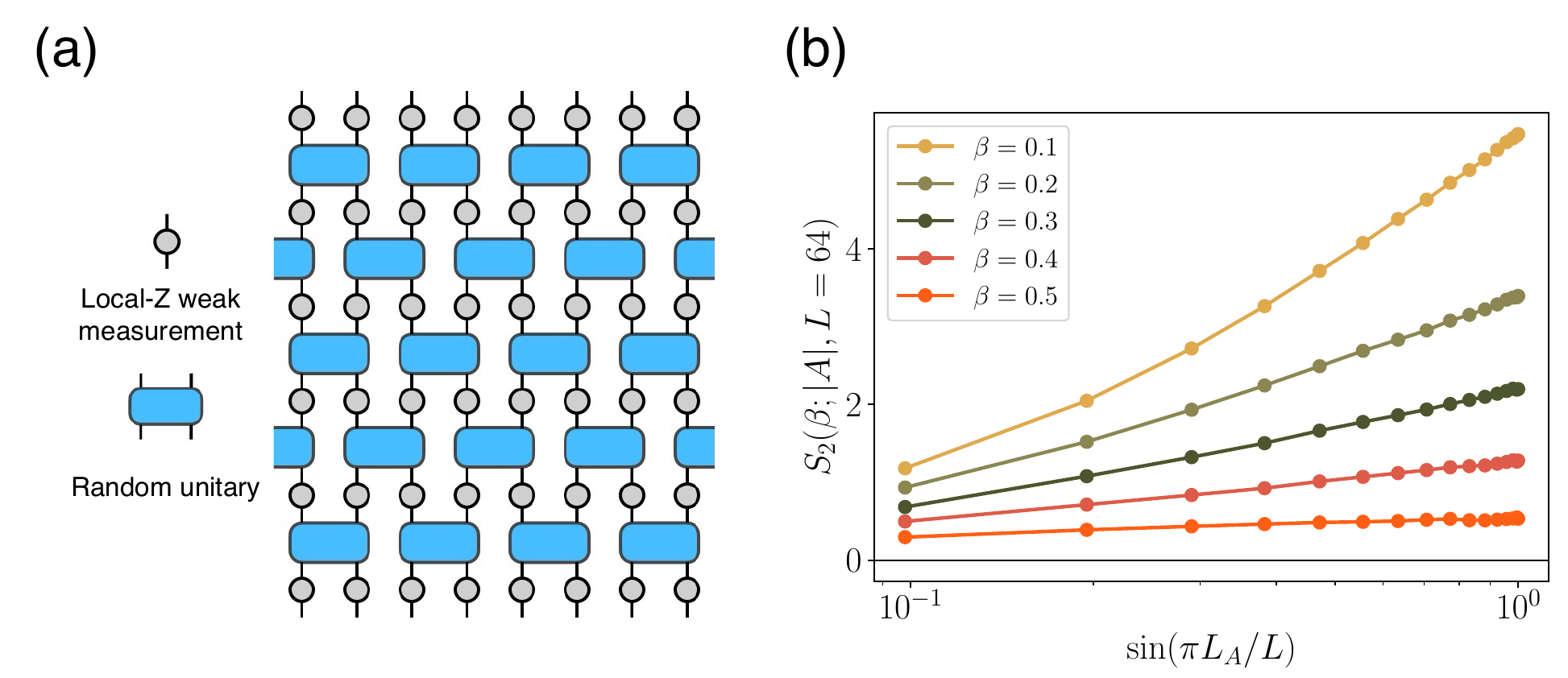}
    \caption{Unitary free-fermion circuits with random weak measurements. (a) The structure of the circuit.  The weak local-Z measurements are applied to each site parameterized by $\beta$ as in Eq.~\ref{eq:kraus}. (b) The second R\'enyi entanglement entropy $S_2$ for a fixed system size ($L = 64$) as a function of subsystem size $L_A$. We take  $\sin(\pi L_A / L)$ as $x$-axis to account for the periodic boundary conditions. As the weak measurement rate varies, the scaling $S_2$ undergoes a transition from logarithmic scaling $\log(\sin(\pi L_A / L))$ at small $\beta$ to a finite constant at large $\beta$.
}
    \label{fig:mi_system_size_weak}
\end{figure}

We further investigate the transition of magic in hybrid circuits with weak measurements, see Fig.~\ref{fig:mi_system_size_weak}(a) for a schematic structure. Weak measurements represent a generalization of projective measurements that partially collapse the quantum state while retaining some coherence. In our setup, we implement local weak measurements in the $Z$-basis, described by the non-unitary operator:
\begin{equation}
\label{eq:kraus}
K_{j,\pm \beta} = \frac{\exp\left(\pm\beta  Z_j\right)}{\sqrt{2\cosh 2\beta}} = \frac{\exp\left(\pm \beta (2c_j^\dagger c_j - 1)\right)}{\sqrt{2\cosh 2\beta}},
\end{equation}
where the sign in front of $\beta$ denotes the measurement outcome on the $j$-th qubit and $\beta \geq 0$ parameterizes the measurement strength. The measurement statistics and post-measurement state evolution follow:
\begin{equation}
\begin{aligned}
P_{\pm} &= \text{Tr}\left[\rho(t) (K_{j,\pm \beta})^2\right], \\
\rho &\to \frac{K_{j,\pm \beta} \rho K_{j,\pm \beta}}{P_{\pm}}.
\end{aligned}
\end{equation}

The measurement strength $\beta$ is the tuning parameter of the entanglement phase transition: we have logarithmic scaling and area-law entanglement respectively on the two sides the transition point. 

The properties of SRE and SMI are in line with those in the circuit with projective measurements. Specifically, we see a change from an area-law to logarithmic scaling as we tune $\beta$ across the same critical point (Fig.~\ref{fig:mi_system_size_weak}(b)(c)), despite that SRE remains extensive across all measurement strengths (Fig.~\ref{fig:mi_system_size_weak}(a)). 

\begin{figure}
    \centering
    \includegraphics[width=\linewidth]{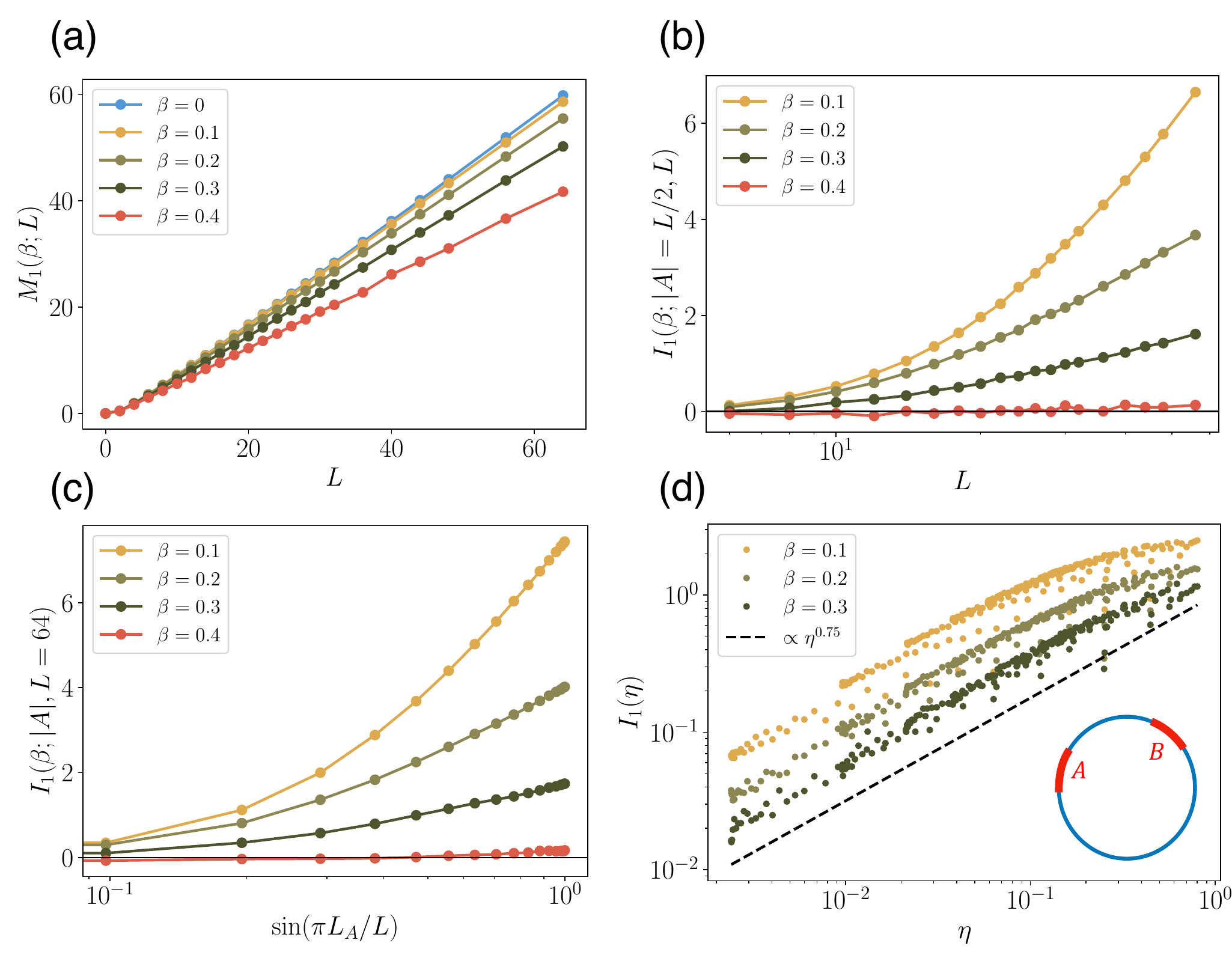}
    \caption{The SRE and SMI of the steady state generated by the free fermion circuits with weak measurements. (a) The total SRE for different measurement strengths $\beta$, all of which exhibit a linear scaling with system size, $M\sim L$. (b) The SMI with a fixed subsystem fraction ($|A|/L = 1/2$) as a function of system size $L$. Without measurements ($\beta = 0$), the BSMI follows a volume-law scaling (not shown).  With small but nonzero $\beta$, BSMI exhibts a logarithmic scaling $I\sim \log L$. Above the critical measurement rate, the BSMI transitions to an area-law scaling $I\sim {\rm const}$. (c) The BSMI for a fixed system size($L = 64$) as a function of subsystem size $L_A$. We take $\sin(\pi L_A / L)$ as the $x$-axis to account for the periodic boundary condition. As the weak measurement strength varies, the BSMI undergoes a transition from logarithmic scaling, $I\sim\log(\sin(\pi L_A / L))$, to area-law scaling above critical point. (d) The SMI for two disjoint regions $A$ and $B$ on the ring. The data points are found to collapse to a single curve depending only on the cross ratio $\eta$.}
    \label{fig:mi_system_size_weak}
\end{figure}

Finally, we study the SMI of two subregions $A = [x_1, x_2] $ and $ B = [x_3 , x_4]$ on a periodic chain. In the critical phase, the mutual information of entanglement demonstrates the scale invariance of the system and depends only on the cross ratio defined as~\cite{chen2020conformalfreefermion}
\begin{equation}
    \eta=\frac{x_{12} x_{34}}{x_{13}x_{24}},~x_{ij}=\sin({\frac{\pi}{L}|x_i-x_j|}).
\end{equation}
We test the scaling of the SMI in the same two-subregion setup. There is one caveat: we note that locality in the Pauli basis and free fermion basis are not identical for two \textit{disjoint} regions. Under a Jordan-Wigner transformation, Pauli operators supported in disjoint regions may map to fermionic strings spanning the intermediate region, and vice versa. Considering the fact that the criticality is observed in the fermion basis, we therefore compute a variant of SRE by sampling fermionic operators supported only in $A$ and $B$ for the disjoint regions:
\begin{equation}
    M^f_{\alpha}(\rho_{A\cup B}) = \frac{1}{1-\alpha} \log \bigg[ \frac{\sum_{\gamma_{\mathbf{x}}\in A \cup B} |\mathrm{Tr}(\rho \gamma_{\mathbf{x}})|^{2\alpha}}{\sum_{\gamma_{\mathbf{x}}\in A \cup B} |\mathrm{Tr}(\rho \gamma_{\mathbf{x}})|^{2}} \bigg].
\end{equation}

The result is presented in Fig.~\ref{fig:mi_system_size_weak} (d), which demonstrates that the SMI depends solely on the cross ratio. When the subsystems $A$ and $B$ are taken to be small and far apart ($\eta\to 0$), the SMI scales as a power of cross ratio: $\eta^{\alpha}$. We observe that $I_{1}(A,B)\propto \eta^{0.75}$ and $I_{2}(A,B)\propto \eta^{0.93}$. It indicates that the magic also captures the universal data of the critical phase. 

\subsection{Dynamics}
\label{sec:dyn}
Beyond the steady state behaviors, we also investigate the dynamics and relaxations of the SRE and SMI in random free fermion circuits.

Previous studies on Haar random circuits have shown that magic grows much more rapidly than the entanglement. For example, the magic can reach the steady state value (within a fixed error $\epsilon$) in a depth that scales as $ \mathcal{O}(\log L) $\cite{turkeshi2024magicspreadingrandomquantum}. Specifically, numerical evidence supports an exponential approach to the steady state for a generalized stabilizer entropy (which includes the SRE) in a form $ \Delta M(t ) \equiv|M(t) - M(\infty)| \sim L \exp(-a t) $, where $ a $ is a size-independent constant. As a comparison, entanglement entropy can only grow  linearly in time for local circuits and saturates over a timescale of $ \mathcal{O}(L) $.

We investigate the growth and saturation of magic in random free fermion circuit and find it significantly slower for both purely unitary dynamics and the critical phase under measurements. 

\begin{figure}
    \centering
    \includegraphics[width=\linewidth]{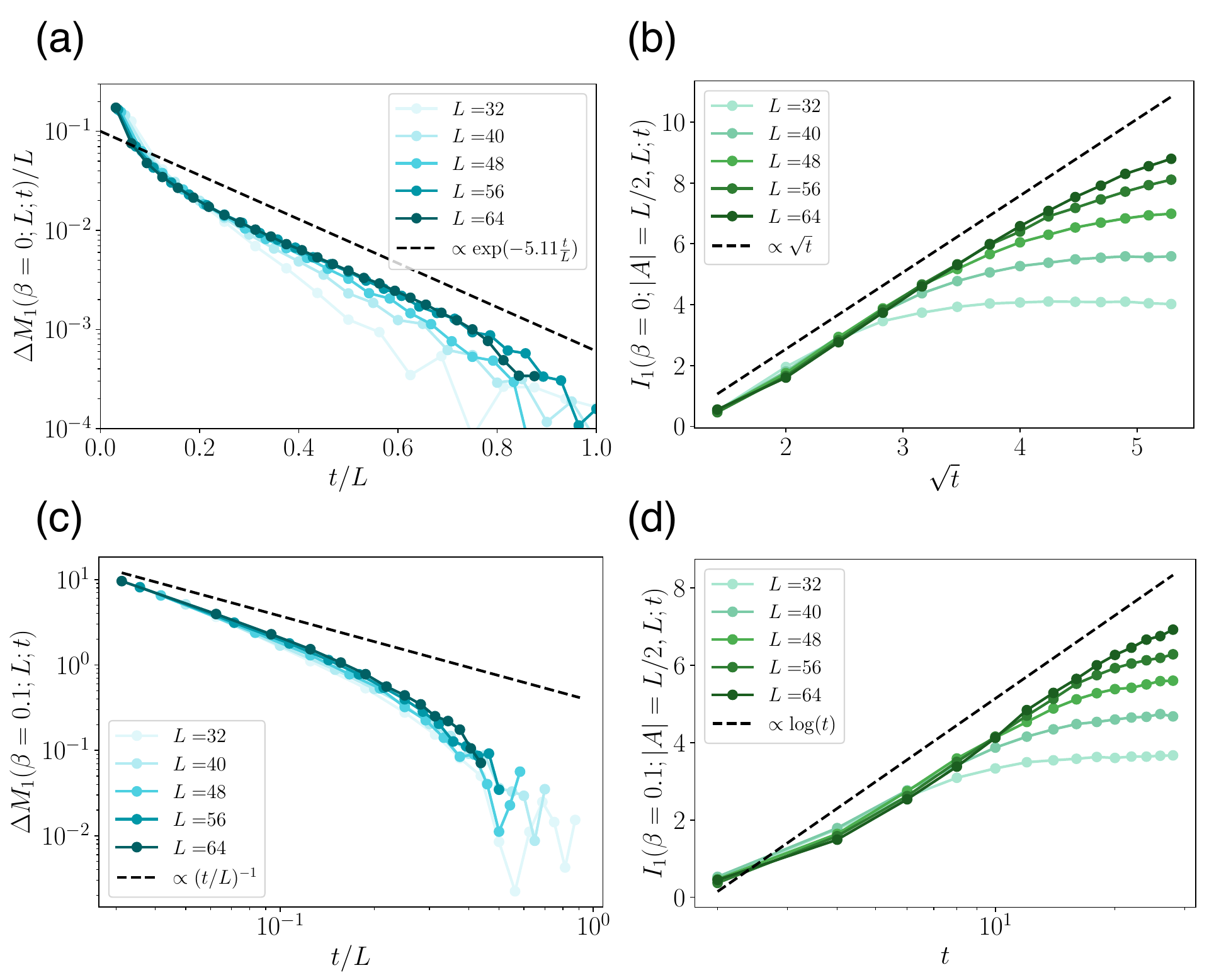}
    \caption{The dynamics of SRE and SMI under purely unitary free fermion evolution and non-unitary evolution with fixed weak measurement strength $\beta=0.1$. (a) The exponential saturation of total SRE under purely unitary free fermion evolution. The deviation $\Delta M_1(t) \equiv M_1(\infty) - M_1(t)$ decays exponentially as $\Delta M_1/L \sim \exp(-5.11t/L)$. Finite-size scaling shows universal collapse, implying a saturation timescale $t^{\rm sat} \sim O(L\log L)$. (b) The sublinear growth of BSMI under purely unitary free fermion evolution. The BSMI exhibits $\sqrt{t}$ temporal scaling, matching the entanglement entropy growth in unitary free-fermion systems with a Bell-pair diffusion picture \cite{chen2020conformalfreefermion}. (c) The saturation of total SRE under non-unitary free fermion evolution with fixed weak measurement strength. The difference between the saturation value of SRE and SRE at time $t$ decays polynomially with time at the early stage of evolution. The curves with different system sizes are found to collapse together when we change the $x$ axis to $t/L$. The saturation time is proportional to the system size $t^{\rm sat}\sim O(L)$. (d) The logarithmic growth of BSMI under non-unitary free fermion evolution with fixed weak measurement strength. The BSMI grows logarithmically with time, in consistent with the growth of EE in non-unitary free-fermion systems \cite{chen2020conformalfreefermion}.}
    \label{fig:sre_dynamics_renyi1}
\end{figure}

We begin with purely unitary dynamics. As shown in Fig.~\ref{fig:sre_dynamics_renyi1} (a), the SRE exhibits a similar exponential decay toward its steady-state value; however, the decay rate is inversely proportional to the system size, as demonstrated by the scaling collapse of the data $\Delta M(t)/L$ as a function of $t/L$. As a result, the saturation time---defined as the time after which the deviation satisfies $|\Delta M_{\alpha}(t)| \le \epsilon$---scales as $\mathcal{O}(L \log L)$, which is substantially longer than that in Haar random circuits. The slow relaxation compared to the Haar circuit can be attributed to the constraints of the free fermion evolution given by Eq.\eqref{Eq:unitary}: it conserves the number of Majorana operators in a Majorana string.  In classical systems such as the symmetric exclusion process, such a constraint generally leads to a diffusive scaling. We hypothesize the lack of diffusive scaling in the relaxation of SRE to a limited simulation time and leave it for future investigation. In addition, the bipartite SMI (See Fig.~\ref{fig:sre_dynamics_renyi1} (b)) exhibits diffusive growth over time.  Such diffusive growth also occurs in the entanglement of the unitary free fermion dynamics \cite{chen2020conformalfreefermion}.

We now turn to the study of non-unitary dynamics. In the critical phase, we find that the BSMI grows logarithmically in time, similar to the entanglement growth in this phase (See Fig.~\ref{fig:sre_dynamics_renyi1} (d)). Additionally, we investigate the relaxation of the SRE for the total wavefunction. As shown in Fig.~\ref{fig:sre_dynamics_renyi1} (c), with weak measurements, $\Delta M(t)$ collapses as a function of $t/L$, indicating that the system takes $\mathcal{O}(L)$ time to saturate to its steady-state value. Specifically, in the early-time regime where $t/L \ll 1$, we observe $\Delta M(t) \sim L/t$ as presented in Fig.~\ref{fig:sre_dynamics_renyi1} (c). In the late-time regime $t/L \gg 1$, we expect $\Delta M(t)$ to decay exponentially as $\exp(-\alpha t/L)$. However, due to the small magnitude of $\Delta M$ and strong fluctuations of logarithmic scale in this regime, we are unable to numerically confirm this behavior. Similar results for the critical phase with projective measurements are presented in the App.~\ref{app:projective_dynamics}. The data for the dynamics with projective measurements do not collapse as well as those with weak measurements, partly because the projective measurements destroy too much magic each time, causing the SRE to fluctuate rapidly as it approaches saturation. 

\section{Conclusion and outlooks}
\label{sec:con}
In this work, we investigate the statistical behaviors of magic in a hybrid free fermion circuit that are known to have an entanglement phase transition from a critical phase to an area-law phase. Our central finding is that although magic, as quantified by the stabilizer R\'enyi Entropy (SRE), remains extensive in both phases, the structure of the magic undergoes a delocalization transition across the critical point, and the dynamics of magic saturation is much slower than that in generic random hybrid circuits. 

We first identify a qualitative change in the structure of magic across the entanglement phase transition. To capture this change, we define the bipartite stabilizer mutual information (BSMI), which isolates the contribution of long-range magic—that is, magic generated by operations that are both highly non-Clifford and long-range entangled when expressed in the Pauli product basis. We show that BSMI scales logarithmically with system size in the critical phase, but remains an 
$O(1)$ constant in the area-law phase—mirroring the scaling behavior of entanglement entropy. The BSMI thus
captures the delocalization transition of magic concurring with the entanglement transition.

Notably, the relation between quantum entanglement and magic remains a compelling open question. In this work, we employ SMI, which in its construction intuitively cancels out local contributions, and show that it successfully echoes the scaling behaviors of entanglement in both phases. However, unlike entanglement and mutual information, SMI can assume either signs. In the models of this work, the sign depends on the R\'enyi index, which requires a proper interpretation. It remains an open question whether other proposed measures of non-local magic, such as those relating to the entanglement spectrum flatness \cite{cao_gravitational_2025}, would exhibit similar scaling behaviors across the phases. 

Regarding the dynamics, we find that the relaxation process leading to the steady states is considerably slower in the free fermion circuits. Specifically, in the purely unitary case, the saturation time scales as $L \ln L$; in the critical phase, it scales linearly $\sim L$ and the SRE collapses as a scaling function of $t/L$. Both are much larger than the  $\log L$ time saturation observed in typical random unitary circuits. In the area-law phase, by contrast, the relaxation is still given by the exponential form without the size-dependent slowdown.

Our investigation demonstrates that the non-local magic structure serves as a valuable probe for universal data from the critical behaviors of hybrid circuits. A remaining question of this work is what aspect of universal data it extracts from the criticality, and how it is related to the computational complexity~\cite{hoshino2025stabilizerrenyientropyconformal,Sierant_2022}. Exploring the behavior of non-local magic in a broader range of quantum critical phases and other non-unitary hybrid dynamics merits additional exploration \cite{Bejan_2024}.  

In the relaxation dynamics of SRE, we observe that magic generation and saturation are substantially slower in the critical free fermion circuits than in Haar random settings. A rudimentary analysis based on operator dynamics in the Majorana basis suggests that the Majorana operator number conservation would give rise to diffusive scaling. Yet we see only a saturation time linear in the system size $L$ rather than a quadratic dependence. A precise interplay of conserved quantities and the observed relaxation time scale requires further investigation, perhaps starting from the simplest unitary dynamics with similar constraints. More broadly, understanding the mechanisms governing the spread and relaxation of magic under general Hamiltonian evolution is an important avenue for further research. 

\section{Acknowledgment}
As we were finalizing our work, we became aware of a recent preprint on arXiv that also explores magic near the measurement-induced phase transition in free fermionic systems \cite{tirrito2025magicphasetransitionsmonitored}.
We thank Hanchen Liu and Zongyue Hou for helpful discussions. This work is supported in part by Grant No. 12375027 from the National Natural Science Foundation of China (Z.-C. Y.). Numerical simulations were performed on the High-performance Computing Platform of Peking University. 

\appendix

\section{Stabilizer Mutual Information of Random States}

\label{app:detail_def_SMI}

\begin{figure*}
    \centering
    \includegraphics[width=\linewidth]{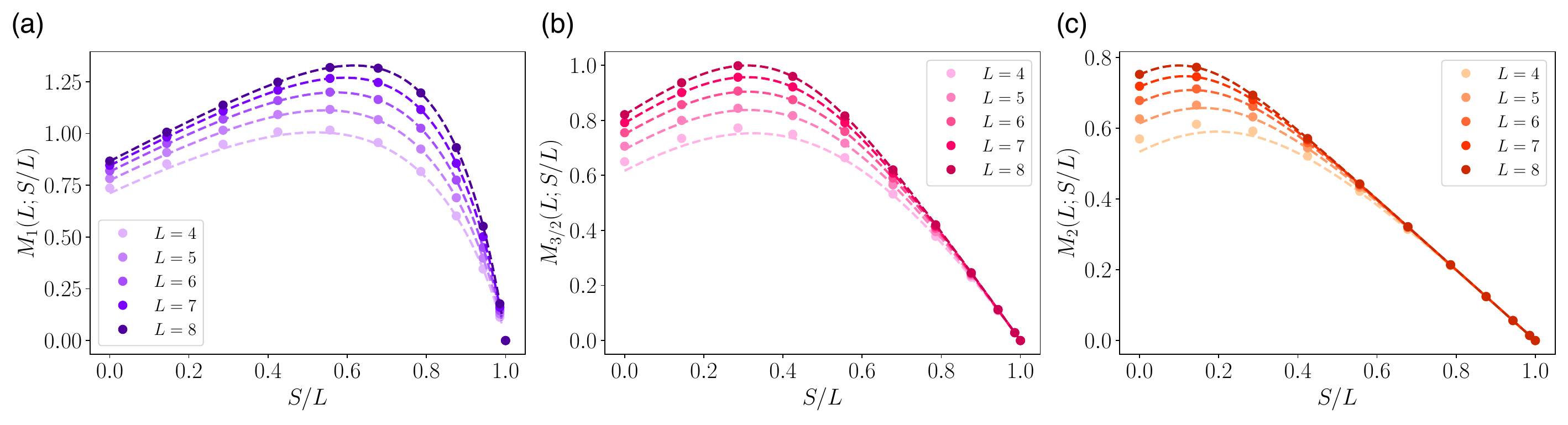}
    \caption{The averaged SRE with R\'{e}nyi index (a) $\alpha=1$ (b) $\alpha=3/2$ (c) $\alpha=2$ for Haar random states. The dotted lines represent the analytical predictions from Eq.~\ref{eq:sre_haar_random_state}, with each line corresponding to a system size of the same color. The data points denote numerical results for Haar random states with system sizes ranging from 4 to 8, averaged over multiple samples. The figure demonstrates that the analytical predictions agree well with the numerical results, confirming their validity in generally random states.}
    \label{fig:sre_ee}
\end{figure*}

To understand the relation between entanglement and magic, we perform a case study of various orders of SRE for Haar random states conditioned on a given purity. In the following, we assume that the density matrix's Pauli spectrum follows a Guassian distribution with constraint. SRE of different R\'enyi indices are computed based on the Guassian assumption, which reveals the relation between entanglement and SRE.  We numerically confirmed these relations by getting agreements with the analytic results in Fig.~\ref{fig:sre_ee}. They provide practical guide about how SMI behaves in random states. 

We begin by considering the probability distribution of the Pauli spectrum, defined as
\begin{equation}
    \Pi(x) = \sum_P \delta(x - \mathrm{Tr}(\rho P)).
\end{equation}
Using this probability distribution, we can express the SRE of any order as
\begin{equation}
    M_{\alpha}(\rho) = \frac{1}{1-\alpha} \log \bigg[ \frac{\int {\rm d}x \, \Pi(x) x^{2\alpha}}{\int {\rm d}x \, \Pi(x) x^2} \bigg].
\end{equation}

We sample an $N$-qubit state from the Haar random state ensemble with a log-purity (or the second-R\'enyi entropy) given by $S = -\log \big[ \mathrm{Tr}(\rho^2) \big] = -\log \big[ \frac{\sum_P \mathrm{Tr}^2(\rho P)}{2^N} \big]$. With no prior information about the state, except for the expectation values of specific operators, such as the identity operator, which satisfies $\mathrm{Tr}(\rho I) = 1$, we assume that the density matrix components in the Pauli basis are randomly distributed with Gaussian randomness~\cite{turkeshi2025paulispectrum}. This leads to the probability distribution of the Pauli spectrum being modeled as
\begin{equation}
    \Pi(x) = A e^{-x^2 / 2\sigma} + b_1 \delta(x - 1) + (b - b_1) \delta(x + 1),
\end{equation}
where $b$ is the total number of predetermined operators, typically equal to one. The coefficients $A$ and $\sigma$ are determined by the total dimension of the Pauli operator space and the purity of the state,
\begin{equation}
    \int {\rm d}x \, \Pi(x) = 4^N, \quad \int {\rm d}x \, \Pi(x) x^2 = 2^{N-S}.
\end{equation}
By solving these conditions, we obtain the explicit form of the probability distribution $\Pi(x)$:
\begin{equation}
    \Pi(x) = \frac{4^N - b}{\sqrt{2\pi \sigma}} e^{-x^2 / 2\sigma} + b_1 \delta(x - 1) + (b - b_1) \delta(x + 1),
\end{equation}
where the variance $\sigma$ is given by $\sigma = \frac{2^{N-S} - b}{4^N - b}$.

With the probability distribution $\Pi(x)$, we can compute the SRE of different orders:
\begin{equation}
\label{eq:sre_haar_random_state}
    M_\alpha = \frac{1}{1-\alpha} \log \bigg\{ 2^{-(N-S)} \bigg[ \frac{2^{\alpha} \Gamma(\alpha + 1/2)}{\sqrt{\pi}} \frac{(2^{N-S} - b)^{\alpha}}{(4^N - b)^{\alpha - 1}} + b \bigg] \bigg\}.
\end{equation}
Taking the limit $\alpha \to 1$, we obtain the expression for $M_1$:
\begin{equation}
    M_1 = \big(2^{-(N-S)} b - 1 \big) \bigg( \log \bigg[ \frac{2^{N-S} - b}{4^N - b} \bigg] + \frac{2 - \gamma}{\ln 2} - 1 \bigg).
\end{equation}
Here, $\gamma$ denotes the Euler-Mascheroni constant, defined as, $\gamma = \lim_{n\to\infty} \bigg( \sum_{k=1}^{n} \frac{1}{k} - \ln n \bigg) \approx 0.577$.

In the thermodynamic limit, as $N \to \infty$, if the log-purity $S$ is not comparable to $N$ (i.e., $N-S > \mathcal{O}(1) \sim \log b$), the asymptotic leading term of $M_{\alpha}$ is given by:
\begin{equation}
    M_\alpha \approx \frac{1}{1-\alpha} \log \bigg[ \frac{2^{\alpha} \Gamma(\alpha + 1/2)}{\sqrt{\pi}} 2^{(1-\alpha)(N+S)} + b 2^{-(N-S)} \bigg].
\end{equation}

Interestingly, we find that the scaling behavior of the leading term is closely related to the Rényi index $\alpha$:

(1) For $\alpha \leq 1$, the SRE scales as $M_{\alpha} \approx N+S$, increasing with increasing purity.

(2) For $\alpha \geq 2$, the SRE scales as $M_{\alpha} \approx (N-S)/(\alpha-1)$, decreasing with increasing purity.

(3) For $1 < \alpha < 2$, there exists a critical entropy density at which the scaling of the SRE undergoes a transition. Specifically: When $S/N < (2-\alpha)/\alpha$, the SRE follows $M_{\alpha} \approx N+S$; When $S/N > (2-\alpha)/\alpha$, the SRE instead follows $M_{\alpha} \approx (N-S)/(\alpha-1)$.

This behavior highlights a fundamental shift in how SRE scales with purity depending on the value of the Rényi index $\alpha$.

To validate our assumption, we perform numerical experiments to compute the SRE for Haar-random mixed states. Starting with a mixed initial state of a given purity, we evolve the system using a global Haar-random unitary and evaluate the SRE for different orders. After averaging over multiple samples, we obtain numerical results for the SRE up to $N = 10$ qubits. The results, presented in Fig.~\ref{fig:sre_ee}, demonstrate that our analytical predictions align well with numerical simulations, confirming the validity of our approach in realistic scenarios.

\section{Numerical results for order-2 stabilizer R\'{e}nyi entropy}

\label{app:sre_renyi2}

In this appendix, we present numerical results for the second-order SRE ($M_2$) and the corresponding SMI ($I_2$), further supporting the connection between entanglement and magic discussed in Sec.~\ref{sec:sre_smi}. 

\begin{figure}
    \centering
    \includegraphics[width=\linewidth]{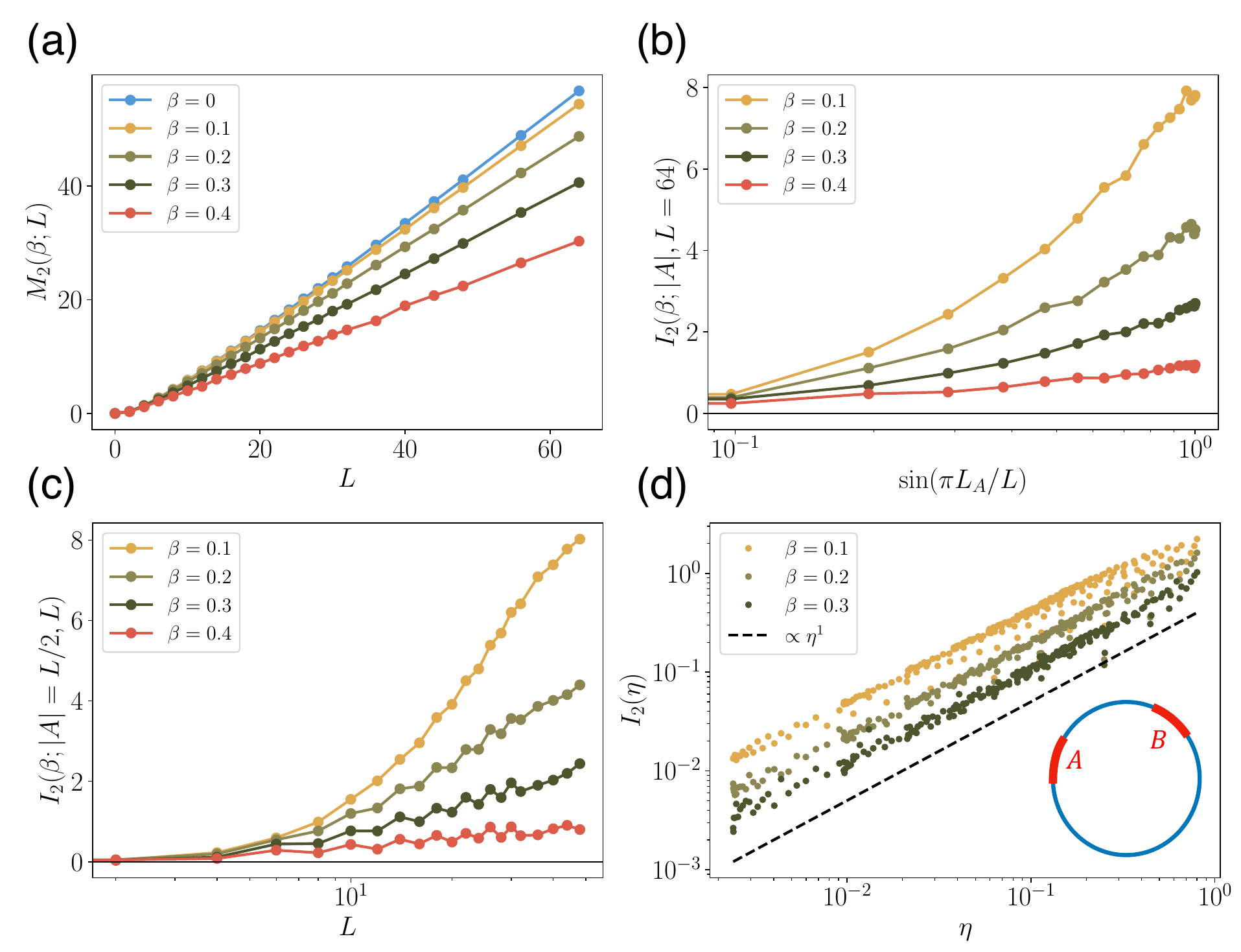}
    \caption{(a) The total $M_2$ of steady states generated by circuits with different weak measurement strength $\beta$. (b) The BSMI with a fixed subsystem fraction, $|A|/L = 1/2$, as a function of system size $L$. (c) The BSMI for a fixed system size, $L = 64$, as a function of subsystem size $L_A$. (d)The SMI for two disjoint regions $A$ and $B$ on the ring. }
    \label{fig:SRE_weak_renyi2}
\end{figure}

\begin{figure}
    \centering
    \includegraphics[width=\linewidth]{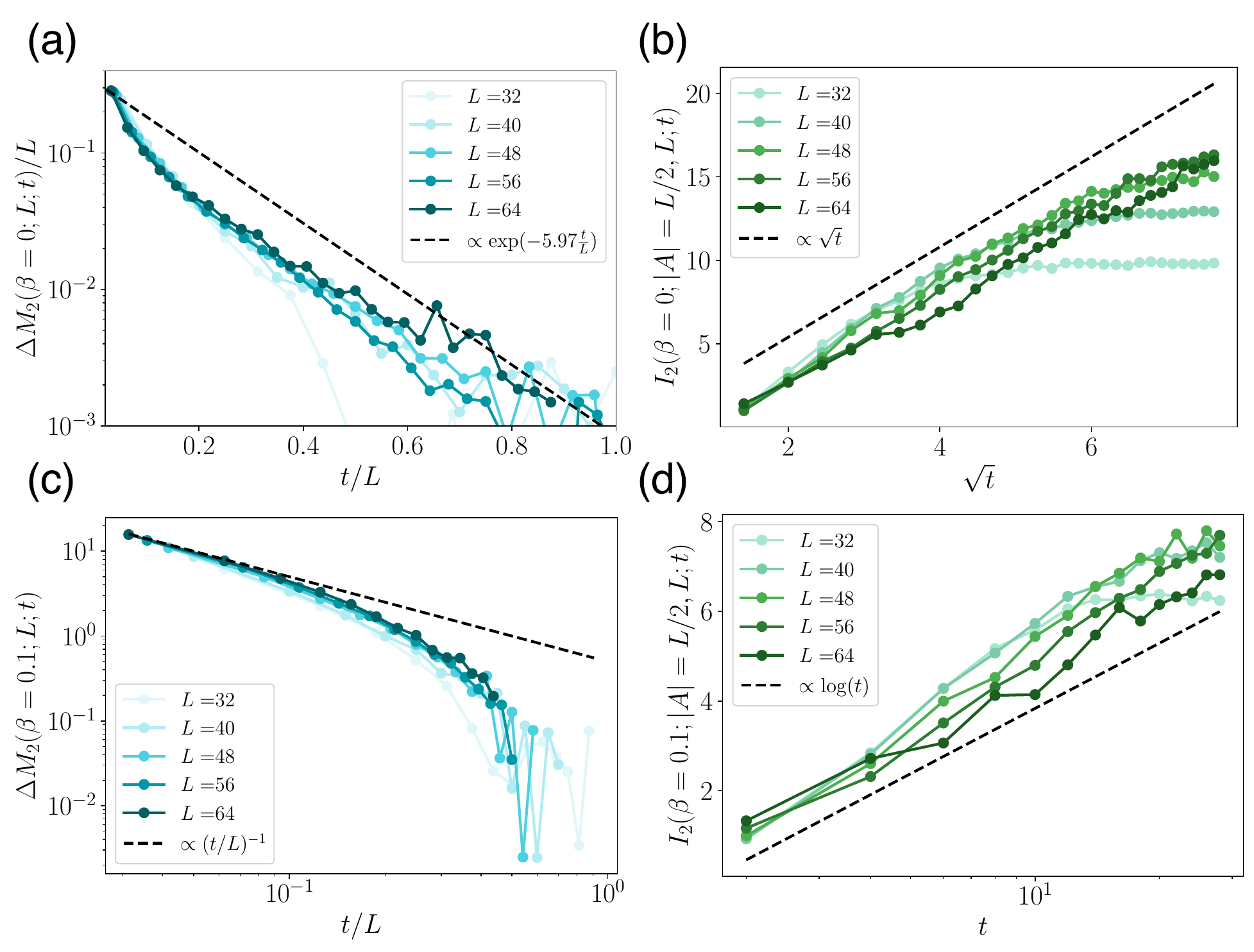}
    \caption{(a) Unitary dynamics of $M_2$. (b) Unitary dynamics of $I_2$. (c) Non-unitary dynamics of $M_2$ with weak measurements. (b) Non-unitary dynamics of $I_2$ with weak measurements.}
    \label{fig:SRE_dynamics_renyi2}
\end{figure}

Fig.~\ref{fig:SRE_weak_renyi2} shows $M_2$ and $I_2$ for steady states of hybrid free-fermion circuits with weak measurements.  
Their phase-dependent scaling mirrors the first-order behavior presented in Fig.~\ref{fig:mi_system_size_weak}. Similarly, Fig.~\ref{fig:SRE_dynamics_renyi2} demonstrates that the dynamical scaling of $M_2$ and $I_2$ matches the first-order case, as discussed in Sec.~\ref{sec:dyn}. These results confirm that our SMI definition in Eq.~\ref{eq:SMI} yields consistent SRE and SMI behavior across different R\'{e}nyi indices.  

\section{Numerical results for the non-unitary dynamics of SRE and SMI with projective measurements}

\label{app:projective_dynamics}

\begin{figure}[!h]
    \centering
    \includegraphics[width=\linewidth]{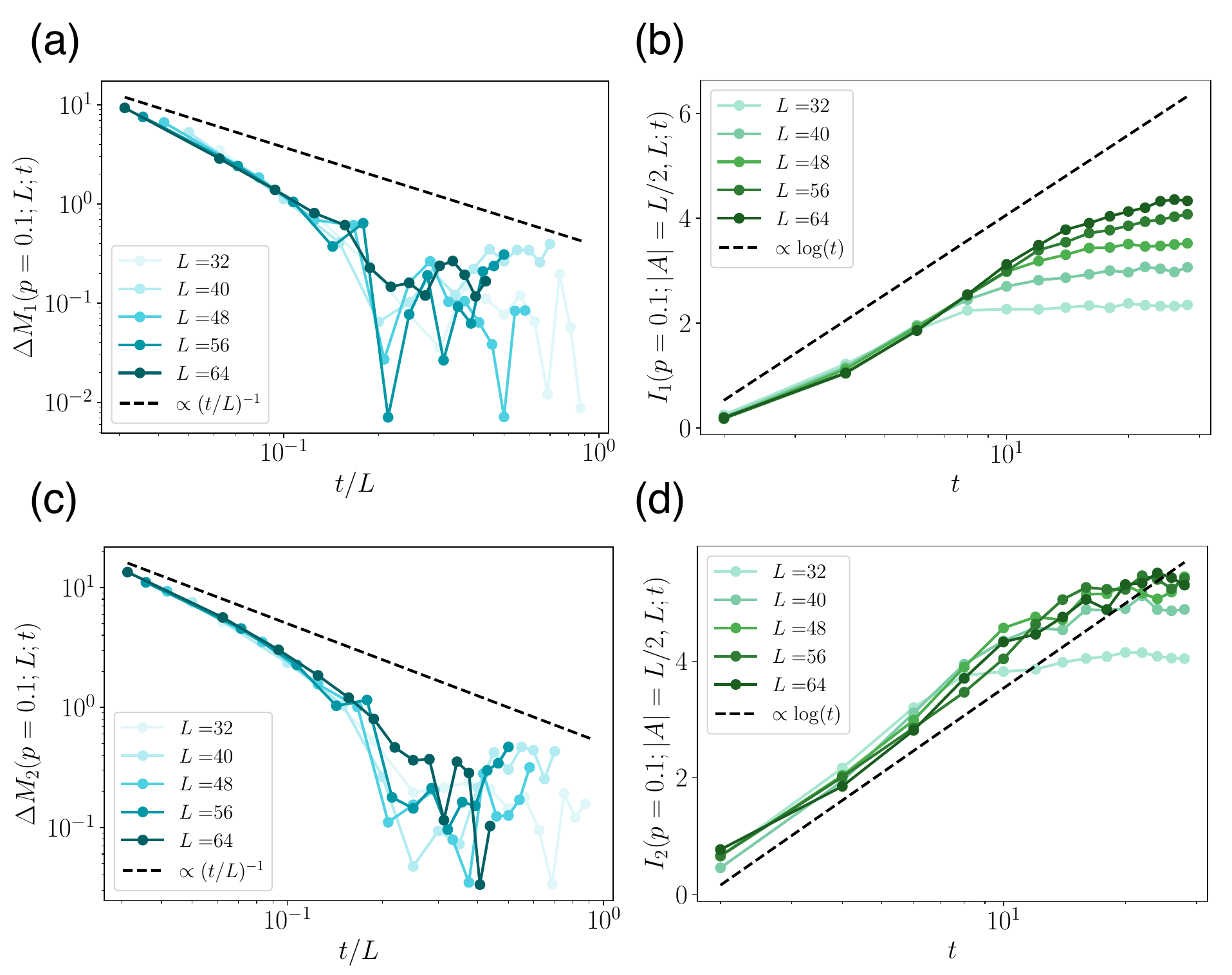}
    \caption{(a) Non-unitary dynamics of $M_1$ with projective measurements. (b) Non-unitary dynamics of $I_1$ with projective measurements.
    (c) Non-unitary dynamics of $M_2$ with projective measurements. (b) Non-unitary dynamics of $I_1$ with projective measurements.}
    \label{fig:sre_dynamics_projective}
\end{figure}

In this appendix, we will show numerical results for the dynamics of SRE and SMI in hybrid circuits with projective measurements. The results are presented in Fig.~\ref{fig:sre_dynamics_projective}. Projective measurements significantly suppress quantum magic, leading to enhanced fluctuations in the dynamics. However, when the axes are rescaled similarly to the weak measurement case in Sec.~\ref{sec:dyn}, the data collapse exhibits a nearly identical structure. This suggests that projective and weak measurements can induce qualitatively similar effects on the dynamics of magic. 

\bibliography{apssamp}

\end{document}